\documentclass[sn-mathphys-num]{sn-jnl}

\usepackage{graphicx}%
\usepackage{multirow}%
\usepackage{amsmath,amssymb,amsfonts}%
\usepackage{amsthm}%
\usepackage{mathrsfs}%
\usepackage[title]{appendix}%
\usepackage{xcolor}%
\usepackage{textcomp}%
\usepackage{manyfoot}%
\usepackage{booktabs}%
\usepackage{algorithm}%
\usepackage{algorithmicx}%
\usepackage{algpseudocode}%
\usepackage{listings}%

\theoremstyle{thmstyleone}%
\theoremstyle{thmstyletwo}%

\theoremstyle{thmstylethree}%

\renewcommand{\orcidlogo}{%
  \includegraphics[width=10pt]{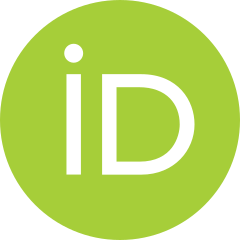}%
}
\renewcommand{\orcid}[1]{\href{https://orcid.org/#1}{\orcidlogo}}

\begin{document}

\title[Exploring Large Quantities of Secondary Data from High-Resolution Synchrotron X-ray Computed Tomography Scans Using AccuStripes]{Exploring Large Quantities of Secondary Data from High-Resolution Synchrotron X-ray Computed Tomography Scans Using AccuStripes}

\author*[1]{\fnm{Anja} \sur{Heim} \orcid{0000-0002-3670-5403}}\email{anja.heim@iis.fraunhofer.de}
\author[1]{\fnm{Thomas} \sur{Lang} \orcid{0000-0001-5939-3919}}\email{thomas.lang@iis.fraunhofer.de}
\author[1,2]{\fnm{Christoph} \sur{Heinzl} \orcid{0000-0002-3173-8871}}\email{christoph.heinzl@uni-passau.de}

\affil[1]{\orgdiv{Division Development Center X-ray Technology}, \orgname{Fraunhofer Institute of Integrated Circuits IIS}, \orgaddress{\street{Flugplatzstr. 75}, \city{F\"{u}rth}, \postcode{90768}, \state{Bavaria}, \country{Germany}}}

\affil[2]{\orgname{University of Passau}, \orgaddress{\street{Innstra\ss{}e 43}, \city{Passau}, \postcode{94032}, \state{Bavaria}, \country{Germany}}}

\abstract{%
The analysis of secondary quantitative data extracted from high-resolution synchrotron X-ray computed tomography scans represents a significant challenge for users. While a number of methods have been introduced for processing large three-dimensional images in order to generate secondary data, there are only a few techniques available for simple and intuitive visualization of such data in their entirety. 
This work employs the AccuStripes visualization technique for that purpose, which enables the visual analysis of secondary data represented by an ensemble of univariate distributions. It supports different schemes for adaptive histogram binnings in combination with several ways of rendering aggregated data and it allows the interactive selection of optimal visual representations depending on the data and the use case. We demonstrate the usability of AccuStripes on a high-resolution synchrotron scan of a particle-reinforced metal matrix composite sample, containing more than 20 million particles. Through AccuStripes, detailed insights are facilitated into distributions of derived particle characteristics of the entire sample. Furthermore, research questions such as how the overall shape of the particles is or how homogeneously they are distributed across the sample can be answered.%
}


\maketitle

\section{Introduction} 
The demand for high quality non-destructive investigations of samples in various fields is constantly increasing, especially in the domain of materials science. Here, the analysis of microstructures is of utmost importance, since these are often directly related to the mechanical, electrical or thermodynamic properties of a material. Synchrotron X-ray computed tomography (SCT) uses the unique properties of synchrotron radiation to facilitate high-resolution examinations of samples with hardly any artifacts. 

A holistic analysis of samples not only requires the analysis of the primary volumetric data~\cite{heinzl_visualization_2021}. For a quantitative inspection of such samples, it is necessary to derive secondary data from the scan, i.e., auxiliary information that is directly related to the microstructure's geometry and properties. Secondary data typically lists derived properties, such as the volume, surface, extent, or shape factor, describing each individual feature of interest, e.g., a particle. For a given property, the secondary information thus contains a respective value for every feature present in the sample. The entirety of this property's data across all features can thus be interpreted as a univariate distribution of that characteristic.

In the case of high-resolution synchrotron imaging, the presence of features of interest, such as inclusions, pores, or particles, may be observed with much larger clarity. To quantify these features, they are further analyzed based on their individual properties, such as their volume or sphericity. Depending on the respective application area, the resulting lists of derived data may be extensive, and pose a significant challenge to existing visualization techniques for presenting the data. Samples may contain millions of particles, fibers, or pores to be quantified. For each application area there are a number of properties of interest to inform the decision-making process of experts. It is therefore necessary to develop novel visual metaphors to facilitate the interpretation of the extensive quantity of secondary derived data to enhance its comprehensibility for the end users.

This work demonstrates the applicability of the AccuStripes~\cite{heim_accustripes_2024} visualization technique on secondary data derived from a high-resolution scan of a metal matrix composite sample. This derived secondary information is represented by histograms which approximate univariate distributions. Aggregation through adaptive binning further considers distribution-dependent properties for an optimal exploration of the secondary information.

The remainder of this paper is structured as follows: After reviewing related work in Section~\ref{sec:relatedwork}, we describe a workflow on how to analyze a SCT scan of a particle-reinforced metal matrix sample by means of the AccuStripes method in Section~\ref{sec:methodology}. The results are presented in Section~\ref{sec:results} and a subsequent discussion elaborates our findings. We will close with a brief conclusion.


\section{Related Work}\label{sec:relatedwork}
Synchrotron imaging is performed with the objective of acquiring high quality, high-resolution data, either in a single scan or in a time-resolved setting~\cite{Honkimäki2020}. Consequently, the analysis of the obtained data is often a subsequent and necessary step, requiring intuitive and accessible visualization~\cite{Heinzl2017}.

An example in this regard is presented by Xu et al.~\cite{xu_3d_2019}, who employ high-resolution SCT imaging for observing cracks in metal matrix composite materials, which integrate an aluminum carrier material combined with various reinforcement materials. A crack segmentation is performed, and the cracks are visualized using 3D renderings of the segmentation mask. The visual analysis of secondary derived data is limited to scatter plots, which are used to investigate the correlation between the volume of the crack and the distance to the fracture surface.
An extension of this approach to time-resolved imaging was proposed by Schwartz et al.~\cite{schwartz_real-time_2022}.
Tekawade et al.~\cite{tekawade_real-time_2022} characterized the porosity of SCT samples using machine learning methods. The resulting data provided 3D renderings. However, this approach does not quantify the rendered porous regions. 
Grau et al.~\cite{Grau2010} enable visual analysis of porous structures from XCT images by using illustrative rendering techniques to better perceive the connectivity between pores and their distributed radii.
Dynamic Volume Lines~\cite{Weissenbock2019} enables the comparison of multiple primary XCT scans by linearizing volumetric data and representing them as 1D Hilbert line plots. Additional charts, such as functional boxplots, and interactive exploration, allow for the investigation of local intensity variations.
Zhang et al.~\cite{Zhang2019} present a visualization approach for analyzing bubble-induced attenuation in rock formations. Their method includes bubble classification, a morphology-based alignment strategy, and a registration technique that links bubble shapes to the porous media surface. Comparative visualization techniques allow detailed inspection of bubbles and surrounding structures.
Overall, a relatively small number of studies considered non-primary data for analysis. This information is often employed solely for the purpose of calibrating beamline measurements, such as in small angle scattering and X-ray spectroscopy~\cite{pandolfi_xi-cam_2018}. 

There are few methods that employ derived secondary data for analysis. 
Reh et al.~\cite{Reh2012} present a visualization pipeline that supports the interactive exploration of porosity by combining quantitative porosity determination with visual analysis. A porosity map, complemented by filtering using parallel coordinates, allows users to explore pores based on their local properties.
Amirkhanov et al.~\cite{Amirkhanov2016} present a workflow that automates defect extraction, classification, and visual analysis to investigate damage mechanisms in interrupted in-situ tensile testing. The employed visualization techniques highlight defects in XCT scans, provide an overview of defect distributions, and estimate the final fracture surface.
FiberScout~\cite{Weissenbock2014} is a visualization framework that facilitates the interactive exploration of fiber-reinforced polymers through multiple visualization techniques. It includes parallel coordinates, a scatterplot matrix, and both 2D and 3D views that allow for comprehensive inspection of fiber structures.
In their study of flood-filling behavior in porous media, Piovesan et al.~\cite{piovesan_4d_2020} analyzed secondary data extracted from SCT scans. They performed a segmentation on each time step and computed secondary information including the contact angles between neighboring grains. Andrieux et al.~\cite{andrieux_synthesis_2018} investigated the composition of metal matrix composites consisting of TiC particles within a Ti matrix using X-ray diffraction tomography and the influence of temperature on reaching a phase equilibrium state. The analysis process extracted several properties, such as the mass fraction of TiC particles over an isothermic heat treatment and the distribution of particle diameters. Line plots are employed for the visualization of these properties, thereby showing that the majority of particles exhibit a relatively small size following heat treatment.
In the sector of battery research, Lang et al.~\cite{lang_big_2024} presented an approach that derives secondary data from a high-resolution SCT scan of a battery's microstructure. Derived characteristics include mass percentages of anode, cathode, etc., and proxies for observing the stability of the manufacturing process. The visualization techniques used include rendering annotations to highlight mass percentages, stacked bar charts to express compactness as a ratio of material to background, and line charts to track the stability of the production process across the sample's rotation axis.

Secondary data can be interpreted as statistical data, more precisely a univariate distribution of a property's values.
Thus, visualization techniques specialized for distributions are required to properly display the derived information. 
Blumenschein et al.~\cite{blumenschein_vplots_2020} provide a comprehensive overview of visualization techniques for statistical data. The authors differentiate between histogram representations, shape-based forms and hybrids thereof. Histogram-based plots are well known in all scientific fields and include bar charts in all variants (regular, stacked, grouped, cumulative), where the height (or increase in height) of the plotted bars informs users about the underlying distribution of the visualized data. Shape-based charts, such as line charts, aim directly at displaying the shape of the distribution using a continuous line with or without filled area, where the height or spread (depending on the orientation) provides insights into the shape of the distribution, e.g., highlighting peaks. 

Both histogram- and shape-based techniques require observers to compare the height of each bar or line in order to evaluate the distribution or comparison. Numerous works have studied these techniques for both exploratory and comparative tasks, yet no technique has been found to outperform the other~\cite{heim_accustripes_2024}. Aigner et al.~\cite{aigner_bertin_2011} found that, in particular for comparison tasks, multiple line plots next to or below each other can lead to visual clutter and hinder efficient comparison of multiple distributions. However, for comparison tasks, the arrangement of small charts side by side has been shown to be advantageous over shared-space techniques where multiple lines or bars are plotted on top of each other in the same chart~\cite{javed_graphical_2010}.
Therefore, more sophisticated approaches are required to facilitate the comparison of distributions. Saito et al.~\cite{saito_two-tone_2005} employed two-tone coloring and visualize distributions as filled line charts beneath each other. The CoSi visualization framework~\cite{heim_cosi_2021} represents secondary data distributions as histograms, which are rendered in the form of a color-coded heatmap. However, CoSi is limited to uniform binning, which is suboptimal for visualizing complex distributions, since inappropriate settings of the bin width may yield misleading representations of the distributions.

In conclusion, a multitude of analysis systems have been presented to investigate high-resolution SCT scans. The majority of systems do not visualize derived secondary data at all, or if they do, only basic visualization techniques such as bar charts, or line charts are employed, which are limited in their effectiveness. Secondary data is often represented as distribution, but visualizing these effectively poses significant challenges to the technique used. The large number of data points in SCT scan secondary datasets leads to visual clutter, making analysis difficult. In addition, many existing visualization techniques lack the ability to compare large numbers of distributions.
AccuStripes~\cite{heim_accustripes_2024} provides a solution in that it allows for a flexible comparison of distributions using adaptive binning techniques. In their study Heim et al.~\cite{heim_accustripes_2024} evaluate the AccuStripes visualization technique exclusively on unimodal distributions sampled from a Gaussian distribution. In the following sections, we will elaborate on how AccuStripes can be used to analyze real-world secondary data of high-resolution SCT scans obtained from a particle-reinforced metal matrix composite material.


\section{Methodology}\label{sec:methodology}
\begin{figure} [tb]
 \centering
 \includegraphics[width=\linewidth]{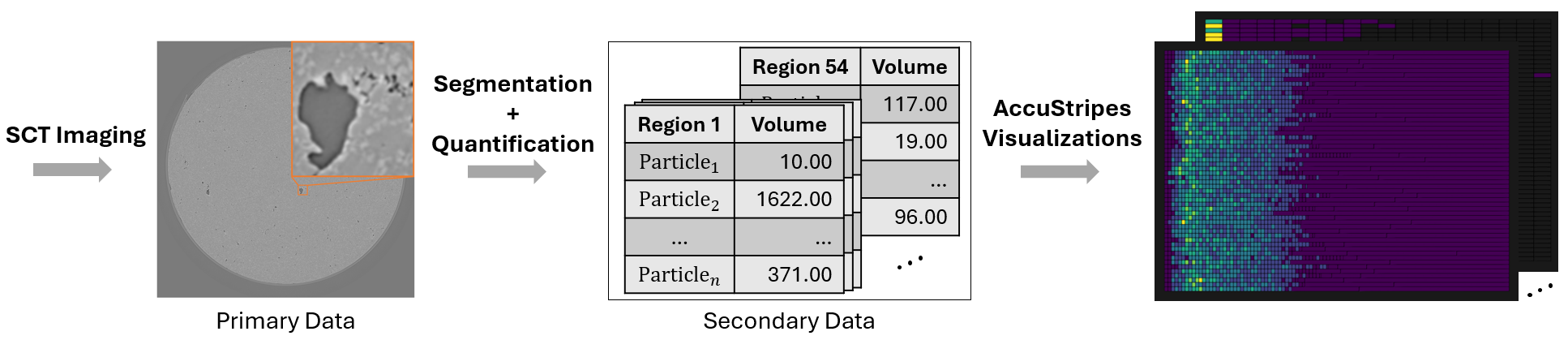}
 \caption{Analysis workflow for SCT scans: Primary data is generated via SCT imaging. Secondary data is derived from the primary data by applying segmentation and quantification processing. The resulting secondary data is then visualized using various AccuStripes visualizations}
 \label{fig:overview}
\end{figure}

In the following paragraphs we describe how large SCT scans of particle-reinforced metal matrix composite materials may be analyzed using an image processing pipeline in combination with the AccuStripes visualization technique. Following an illustration of the quantification process of such SCT scans, a description of visualization techniques that facilitate the interpretation of the quantification will be provided (see Fig.~\ref{fig:overview}).

\subsection{Deriving secondary data from high-resolution SCT scans}
The data investigated in this work are high-resolution synchrotron X-ray computed tomography scans of metal matrix composites. In detail, these materials are highly homogeneous metal alloy samples, which are further reinforced with particles of a different composition to improve the durability of the samples in terms of mechanical or thermodynamical properties. Due to the size of the contained particles, high-resolution scans are required for a non-destructive evaluation of the samples. We consider SCT imaging which possesses a very high flux and an almost parallel beam imaging geometry, which yields (almost) artifact-free scan data. Furthermore, this setup enables the usage of propagation-based phase contrast imaging, which is very sensitive to interfaces between materials. Thus, it enables the detection of fine structures with sharp edges where conventional laboratory XCT acquisition fails to do so. The acquired projection images are subject to artifact corrections and a phase retrieval step before being reconstructed as three-dimensional volumetric data~\cite{lang_multiscale_2023}. Note that a fast high-resolution image acquisition process of industrial samples implies the generation of high numbers of relatively large X-ray projections. Therefore, the reconstructed volumes are typically quite large, reaching Terabytes in size in extreme cases.

The first step towards a quantification of the individual particles is their segmentation. Although, the data quality is very high, we found that a simple thresholding (both with a manually tuned threshold and an automatically derived) tends to oversmooth the result (resulting in an undersegmentation) and prevents obtaining the best possible segmentation. Therefore, a segmentation was obtained using an interactive machine-learning based procedure, proposed by Lang et al.~\cite{lang_geometric_2022}, which exploits local characteristics. This approach makes use of user inputs to train a machine learning model, which is in consequence used to infer a segmentation mask where each voxel is assigned to the probability of being a particle. Even though this technique employs machine learning, this algorithm is specifically designed to handle very large voxel volumes using only a very limited set of interactively provided annotations. In a subsequent post-processing phase, all voxels with a high probability of belonging to particles are retained, resulting in the final segmentation mask.

After the segmentation, most particles can be isolated from each other and secondary information can be derived for each particle. To deal with the considerable number of particles present in high-resolution samples, resulting in a large CSV file of quantified secondary data, we apply a partitioning of the segmented volume into spatial tiles of fixed size. Tiling is performed by subdividing the axis-aligned bounding box of the scan into 54 non-overlapping three-dimensional regions of approximately same size. Each particle is assigned to a tile based on the location of its centroid, creating multiple smaller files, one for each tile. Within each tile a connected components labeling~\cite{beare_optimization_2006} is performed. Then, for each component (corresponding to either a single particle or an inseparable agglomeration of particles), several properties are computed using the Insight Toolkit~\cite{lehmann_label_2007}, including the particle's volume (in voxels), its sphericity (in percent), and the coordinates of its centroid (three coordinates, each in voxels). 
%
This process results in a collection of 54 CSV files for a given property. Each CSV file is associated with a specific spatial tile of the sample and contains the univariate distribution of the corresponding property.

%

\subsection{AccuStripes for particle characteristics}
The AccuStripes~\cite{heim_accustripes_2024} visualization technique enables the representation of distributions in the form of histograms, combining a number of binning techniques with different color composition strategies. Depending on the data and the task at hand, the most suitable visualization for displaying the distributions can be identified interactively. Using the same global limits for all distributions, histograms are computed for each spatial region. A composition strategy is then applied to the histogram to construct the AccuStripes representation. By stacking these representations on top of each other, a comparison across all distributions becomes possible.

In order to compute histograms, it is first necessary to perform a binning. There are three different binning strategies that can be employed: uniform (UB) binning, Bayesian Blocks~\cite{pollack_bayesian_2017} (BB) binning, and Jenks' Natural Breaks~\cite{fisher_grouping_1958} (NB) binning.

UB divides the distribution into bins of equal width with the number of bins being determined by Sturge's rule~\cite{sahann_histogram_2021}, using the maximum number of elements across all parts of the ensemble. A key advantage of UB is its simplicity, requiring only the determination of the global minimum and maximum values without extensive data processing. 
Due to the equal width of the bins, UB is a good binning strategy for approximating uniform or non-tailed densities. However, highly skewed distributions are likely to be misrepresented by UB because sharp peaks tend to be grouped into a single bin without the ability to resolve the shape of the peak more finely.
Nevertheless, binning a heavily tailed distribution with UB offers the benefit of prominently displaying outliers of the data in the sparsely populated bins. Since the bin boundaries are determined independently of the data points, modes or gaps may be divided by these boundaries, potentially leading to misinterpretation by the observer.

In contrast, BB determines bin boundaries based on the data. BB~\cite{pollack_bayesian_2017} places the bin boundaries at locations where the data distribution changes significantly. To find the optimal bin configuration the BB algorithm maximizes a likelihood formulation, merging adjacent bins if this results in an increased overall likelihood. In general, sparsely populated regions or outliers in the distribution are represented in wider bins, while peaks in the data tend to form narrow bins. The ease of identifying gaps, outliers, or peaks, depends on the overall shape of the data. While the representation of symmetric unimodal or uniform distributions may be more effective with UB, BB is an excellent method for resolving gaps in these types of distributions. Therefore, BB is an optimal representation for bimodal distributions, as the sparse region between the two modes can be interpreted as a gap. In skewed distributions, BB provides a finer representation of the peak, capturing the majority of the data more effectively than UB. The localization of outliers in skewed distributions is a challenging process when utilizing BB, as the outliers tend to become obscured within the wider bins of the tail. For multimodal distributions, BB effectively captures gaps and can represent peaks well, especially if they occur in a staircase-like pattern or cover a sufficiently large region. However, in cases where the distribution is characterized by abrupt increases and decreases, the application of BB may result in the partitioning of the peak into multiple narrow bins, which may subsequently lead to misinterpretation by the observer.

Similar to BB, NB~\cite{fisher_grouping_1958} computes the bin boundaries based on the data by performing a clustering procedure analogous to the multi-Otsu scheme~\cite{Liao2001}. This method groups the input data into a number of bins in a process that maximizes inter-class variance while minimizing intra-class variance. Similar to BB, NB is less effective than UB for representing symmetric, unimodal, or uniform distributions since small irregularities in the distribution's symmetry are incorporated into the binning process, potentially distorting its symmetric shape. In contrast, NB is particularly sensitive to peaks but less effective at capturing gaps or outliers. Additionally, NB is also prone to difficulties in visualizing skewed distributions, as it shares the same limitation as UB in dealing with heavily tailed distributions: The majority of the data is grouped into a single bin, while the remaining data is spread across wide, sparsely populated bins. However, NB performs well with non-symmetric bimodal distributions, as it can accurately capture the two distinct peaks. In the context of multimodal distributions, the use of NB is an effective approach for identifying narrow, locally concentrated peaks with a sharp increase in data points.

Following the determination of the binning strategies, AccuStripes offers three composition strategies, categorized as \textit{color only}, \textit{overlay}, and \textit{filled curve}. These strategies determine the presentation of the histograms in accordance with the selected binning. 
In the \textit{color only} composition, all bins are represented by rectangles of the same height. The frequencies of the bins are displayed using a color scheme, with empty bins represented by black and all others represented by colors according to their frequency, mapped in the \textit{viridis} color scheme~\cite{Liu2018}. This composition strategy serves as the foundation for all other composition strategies due to its straightforward interpretability.
The \textit{overlay} and \textit{filled curve} compositions enhance the \textit{color only} strategy by incorporating a curve representing the probability density function, estimated using Kernel Density Estimation~\cite{Silverman1986}. In the \textit{overlay} composition, this curve is depicted as a white line drawn over the colored rectangles. In contrast, the \textit{filled curve} composition crops the rectangles above the curve, filling only the area below it. This approach highlights the frequency values in the bins with a visual representation of the estimated density through the curve's height.

\begin{figure} [tb]
 \centering
 \includegraphics[width=\linewidth]{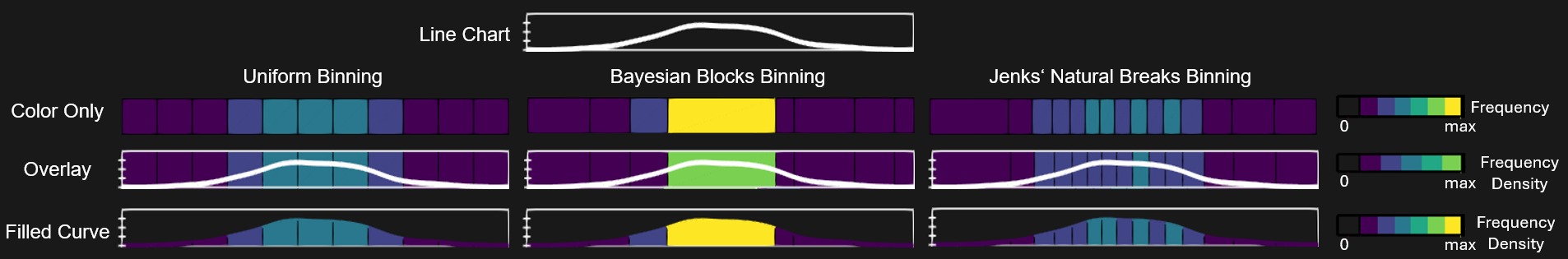}
 \caption{AccuStripes: The Gaussian distribution, shown in the line chart, is visualized by all nine AccuStripes representations. The three compositions are shown from top to bottom; the three possible binning methods are shown from left to right}
 \label{fig:accustripes_overview}
\end{figure}

The combination of a binning technique and a composition strategy results in an AccuStripes representation of a respective distribution of data. For example, Fig.~\ref{fig:accustripes_overview} shows a Gaussian distribution displayed through all nine different AccuStripes representations. For comparison, the individual AccuStripes representations are drawn one below the other, thus allowing the distributions to be compared according to their color pattern. For example, if two distributions are very similar in their shape, such as having a peak of similar height at the same position, the respective bins directly located over each other will have a similar color.

\section{Results \& Discussion}\label{sec:results}
\begin{figure}[tb]
 \centering
 \begin{minipage}{.48\textwidth}
  \centering 
  \includegraphics[width=\linewidth]{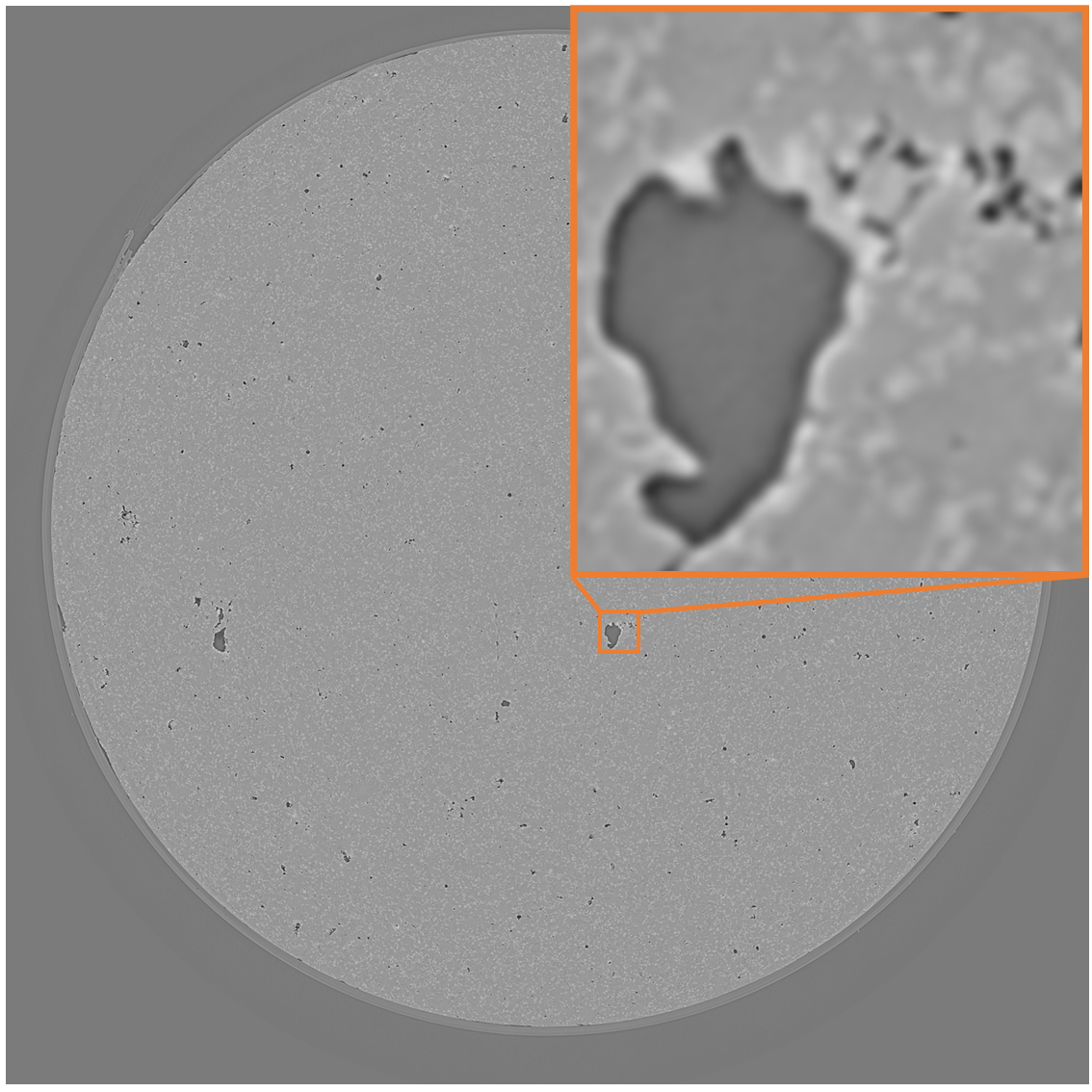}
  \caption{A XY slice of the metal matrix composite considered in this work. The zoomed region shows the contained SiC particles (white) within the metal matrix (gray) and several pores}
  \label{fig:mmcslice}
 \end{minipage}
 \hfill
 \begin{minipage}{.48\textwidth}
  \centering 
  \includegraphics[width=\linewidth]{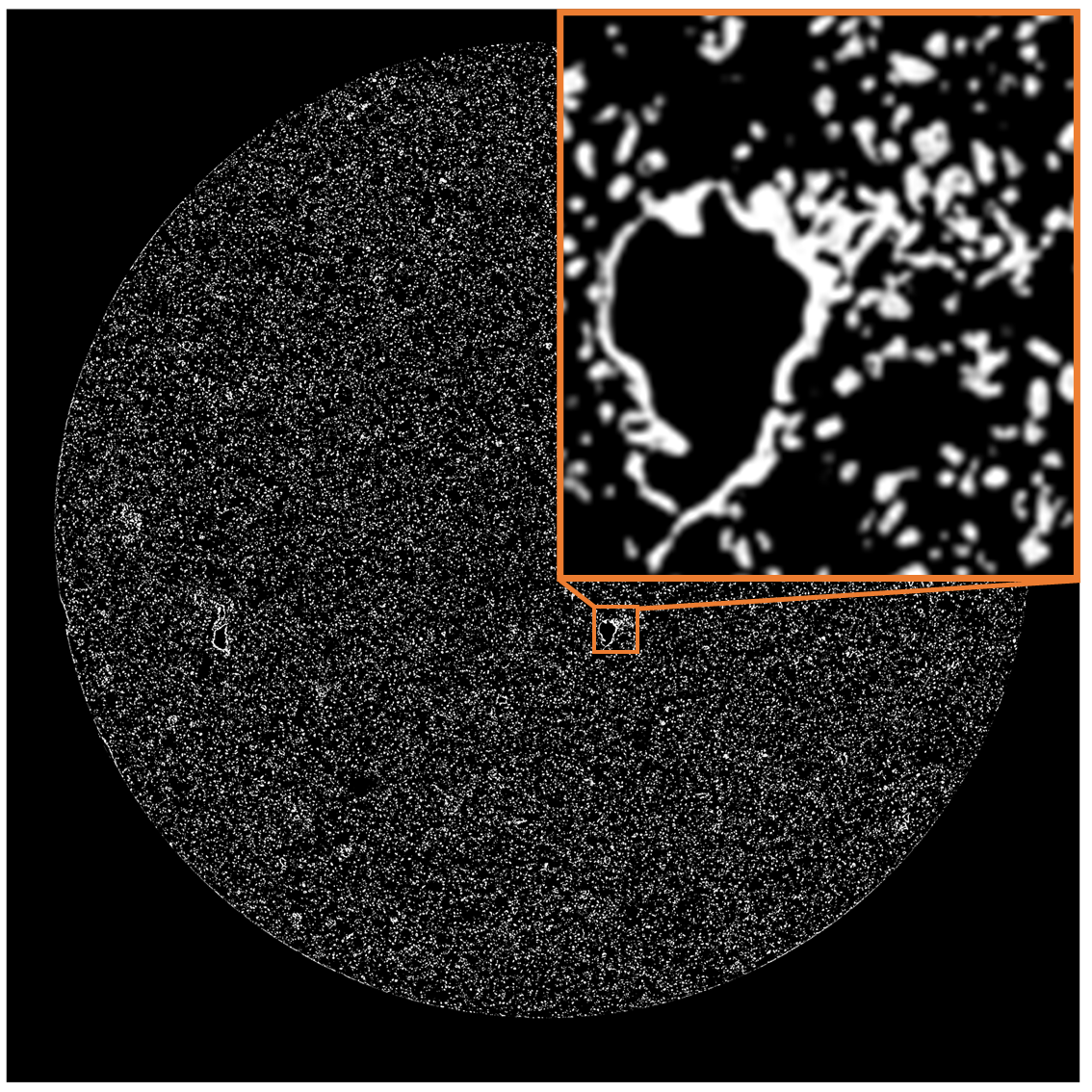}
  \caption{The slice from Fig.~\ref{fig:mmcslice} of the segmentation result. The particles are well segmented and many of them are easily separable in 3D, except for agglomerations around pores}
  \label{fig:mmcseg}
 \end{minipage}
\end{figure}

For the application case presented in this paper, we consider a high-resolution SCT scan of a cylindrical sample of a particle-reinforced metal matrix composite having a diameter of 1~cm and a height of 1~cm. The latter consist of an AlSi matrix material, in which SiC particles are mixed in. The SCT scan was obtained at the BM18~\cite{lang_multiscale_2023} beamline at the European Synchrotron Radiation Facility (ESRF) at a pixel size of 1.2~\textmu{m}. The reconstructed volumetric dataset is stored in 16-bit unsigned integer values and features $8942\times 8942\times 8431$ voxels, which results in a dataset size of 1.35~TB. A slice of this dataset is shown in Fig.~\ref{fig:mmcslice}, illustrating the overall sample and several pores. The zoomed region depicts the individual SiC particles (in white) within the matrix material (gray). From this figure it becomes obvious that a considerable number of particles are dispersed in a relatively homogeneous manner throughout the sample. However, the particles exhibit a tendency to aggregate in clusters around the border regions of the pores. The white pore boundary might also include artifacts from the phase contrast imaging, but it is not possible to distinguish these at this scale.

Using the interactive segmentation technique as prescribed in Section~\ref{sec:methodology} requires selecting individual voxels of particle and non-particle regions and assigning them appropriate labels. A machine learning segmentation model is trained and refined by the interactive selection of additional voxels until the segmentation result is of sufficient quality. Subsequent postprocessing discards all segmented voxels whose probability of being a particle is less than 50 percent, yielding the final segmentation result of which a slice is shown in Fig.~\ref{fig:mmcseg}. Therein, we recon that the particles are segmented well, despite the low visual contrast between them and the matrix material.
The quantification phase processed the resulting volume tile-wise to extract the properties of interest (e.g., volume, sphericity, centroid coordinates) of each particle. The combined information resulted in a CSV file of 1.5~GB in size, containing roughly 20.2 million particles, each attributed with the stated set of properties. 
%
The file was split into five distinct file collections, each corresponding to one of the aforementioned properties. Every collection comprises 54 CSV files, organized according to the spatial tiling of the sample.

\begin{figure}[tb]
 \centering
 \begin{minipage}{.48\textwidth}
  \centering 
  \includegraphics[width=\linewidth]{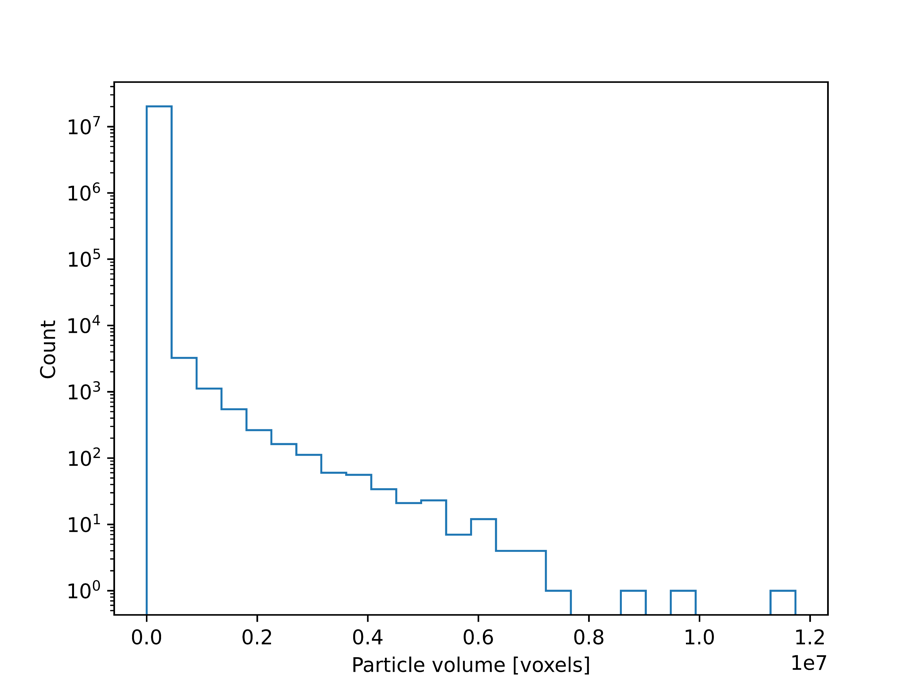}
  \caption{Histogram of the particles' volume showing that most particles have a small volume, while identifying the presence of outliers}
  \label{fig:histvol}
 \end{minipage}
 \hfill
 \begin{minipage}{.48\textwidth}
  \centering 
  \includegraphics[width=\linewidth]{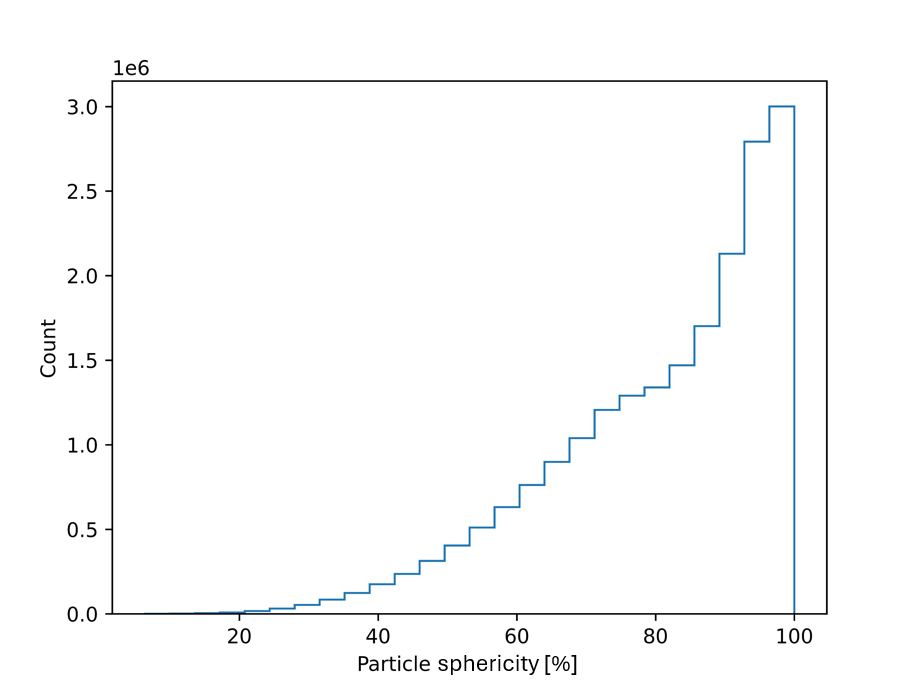}
  \caption{Histogram of the particles' sphericity showing that the majority of particles has a high sphericity}
  \label{fig:histroundness}
 \end{minipage}
\end{figure}

As a first step towards the evaluation of the derived secondary data, we consider histograms of all particles in the sample according to their volume and sphericity. The number of bins was determined by Sturge's rule to be 26. The resulting histograms are displayed in Fig.~\ref{fig:histvol} and Fig.~\ref{fig:histroundness}, respectively. The vast majority of particles show a very small volume, as indicated by the logarithmic scale on the vertical axis of the plot (see Fig.~\ref{fig:histvol}). Furthermore, the distribution is heavily tailed, although being partially masked by the logarithmic axis. It is notable that a small number of components exhibit a very large volume, which does not correspond to individual particles. Rather, these components represent agglomerations of several particles that could not be separated.
Considering the sphericity of the particles, as shown in Fig.~\ref{fig:histroundness}, most particles show high values, i.e., they are close to a mathematically perfect sphere, with only a few showing significant non-sphericity. Compared to the particles' volumes, the trend is inverted, which suggests a correlation between these properties, supporting the intuition that smaller particles are generally more spherical and therefore exhibit higher sphericity. In contrast, components with a high volume are mostly collections of particles clumped together around the shape of a pore. These collections are often not perfectly round, as indicated by the renderings in Fig.~\ref{fig:mmcslice}. Consequently, these cases have a low sphericity.

While these plots provide a general overview of these distributions of the characteristics, there are some drawbacks. Using a logarithmic scale in the histograms brings the benefit of preventing the chart from becoming overly large. However, logarithmic scales may also distort the perception of the data, obscuring long tails in the distribution and making it more challenging to identify outliers or subtle variations. Furthermore, the non-linear relationship between bin height and frequency demands careful interpretation.
%
%
The addition of the complexity of a logarithmic scale on the bins serves to further complicate the process, potentially resulting in a chart that is more difficult and cumbersome to interpret. To minimize complexity, it may be beneficial to crop charts or zoom in on the relevant areas to get insights into the details of interest.

Furthermore, the histograms do not show any spatial relation, e.g., we cannot determine how the particle volume is distributed across the sample or identify regions with a higher concentration of larger particles. While it is possible to visualize each of the 54 CSV files individually through a histogram with a logarithmic scale, the large number of charts and their size would render comparisons of them a cumbersome task. However, this analysis task is facilitated by using AccuStripes, as it allows for a visual examination of multiple distributions.

\begin{figure}[tb]
 \centering
 \includegraphics[width=\textwidth]{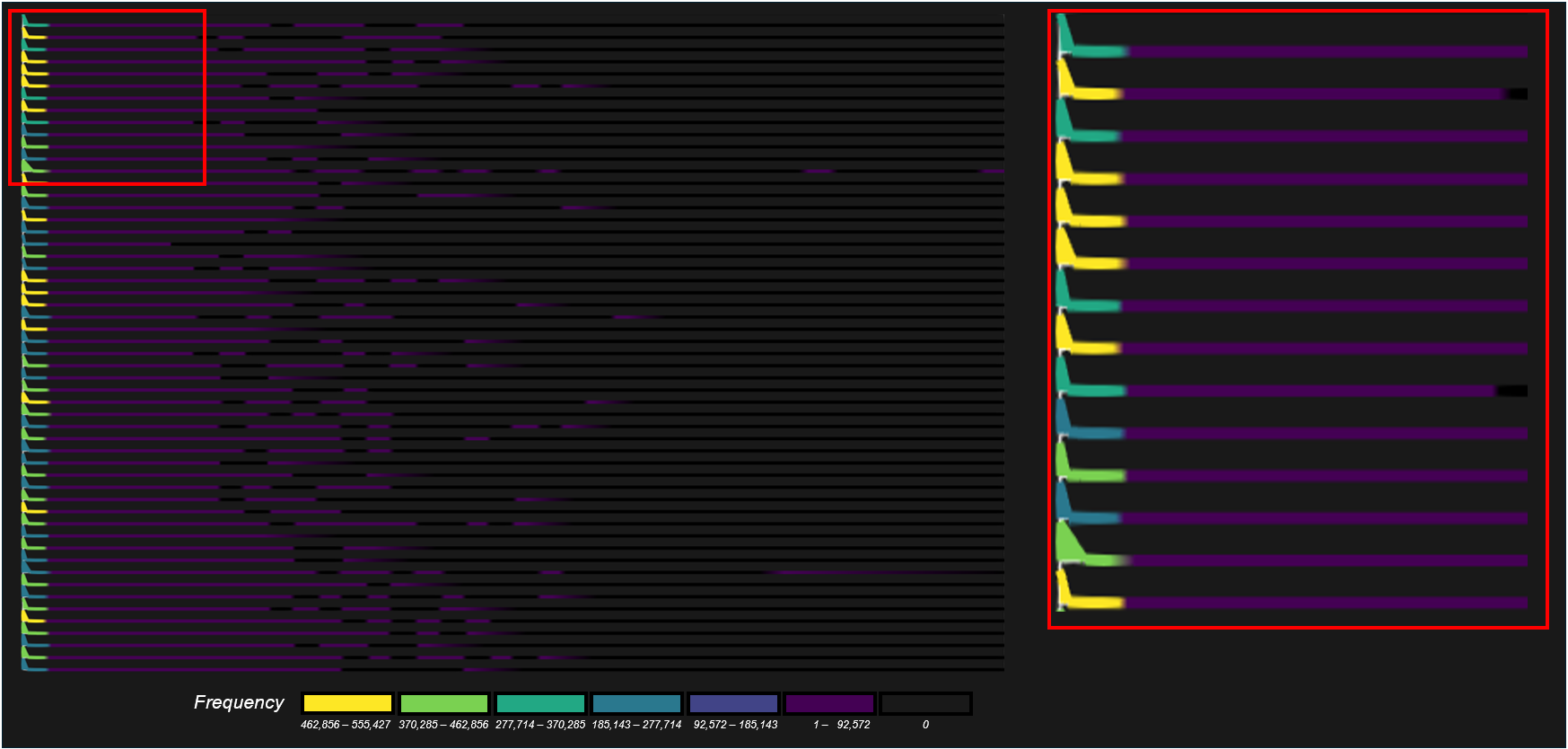}
 \caption{AccuStripes visualization of the particle volume characteristic employing UB and the \textit{filled curve} composition. Due to the shape of the distributions and the 54 non-overlapping spatial volume tiles to be compared, the filled areas are only very small, cf. the detail plot framed in red}
 \label{fig:accustripesColoredLines}
\end{figure}

\begin{figure}[tb]
 \centering 
 \includegraphics[width=\textwidth]{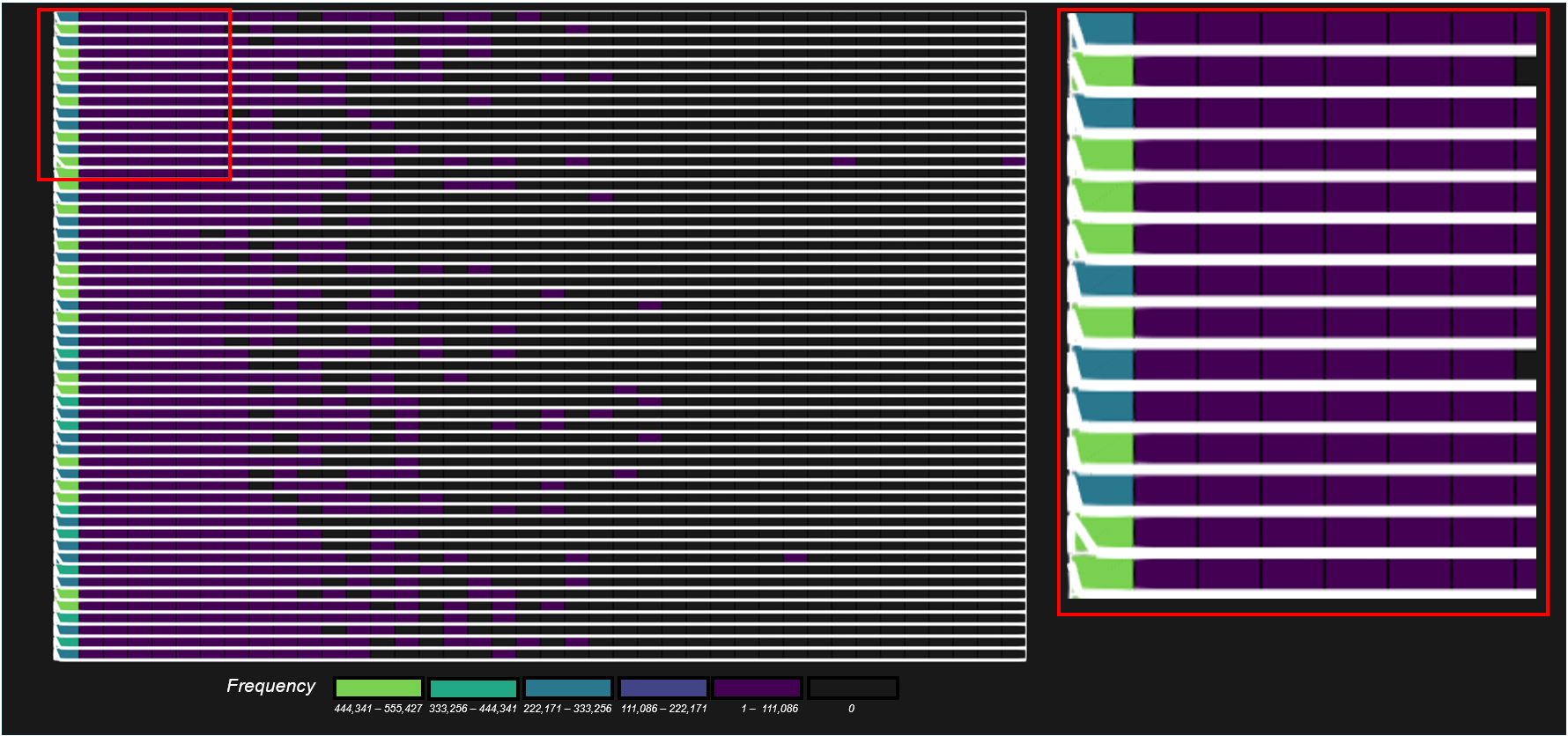}
 \caption{AccuStripes visualization of the particle volume characteristic employing UB and the \textit{overlay} composition. The drawing of the heavily tailed distribution on top of the \textit{color only} bins introduces visual clutter, emphasized by the detail plot framed in red}
 \label{fig:accustripesBinsColoredLines}
\end{figure}

Before discussing the AccuStripes results in detail, it should be noted that this work focuses on the \textit{color only} composition, displaying the bins of the histograms as color-coded rectangles of equal height. While the \textit{overlay} and \textit{filled curve} compositions are available within AccuStripes, these strategies do not improve the interpretability in the context of this work due to the large number of histograms, which need to be stacked on top of each other. Exemplary for the particle volume, Fig.~\ref{fig:accustripesColoredLines} and Fig.~\ref{fig:accustripesBinsColoredLines} show the \textit{filled curve} and \textit{overlay} results, respectively. The considerable number of distributions restricts the height that can be depicted in the visual representations. Moreover, the heavily tailed distributions fill only a tiny area under the curve for the \textit{filled curve} composition (see Fig.~\ref{fig:accustripesColoredLines}). In the \textit{overlay} composition, the additional white line, does not contribute to the interpretability of the visualization (see Fig.~\ref{fig:accustripesBinsColoredLines}). Therefore, this work focuses on the \textit{color only} composition, considering different binning techniques.

We start the analysis with the distribution of the particles' volumes. As described above, the dataset was divided into 54 non-overlapping spatial tiles and each individual tile was binned. 

\begin{figure}[tb]
 \centering
 \includegraphics[width=\linewidth]{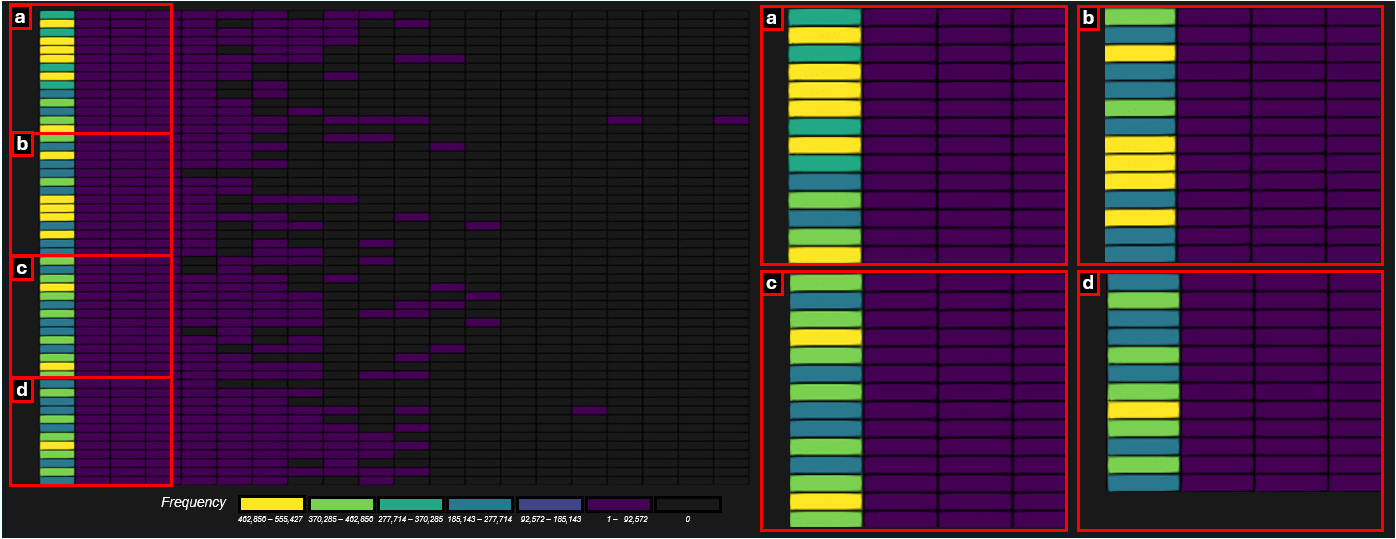}
 \caption{AccuStripes visualization of the particle volume characteristic employing UB and the \textit{color only} composition. The peaks in the distribution over the spatial tiles can be identified, and outliers can be observed with clarity, cf. the detail visualizations (a)-(d). Nevertheless, the peaks remain difficult to analyze in greater detail}
 \label{fig:accustripesVolumeUB}
\end{figure}

In the initial analysis, UB was considered, where the number of bins was determined to be 20 for all distributions as given by Sturge's rule on the maximum number of elements of all rows. The resulting histograms depict the absolute counts/frequency of particles within each bin (see Fig.~\ref{fig:accustripesVolumeUB}) where zoomed regions (a)-(d) show the most interesting regions. As shown in Fig.~\ref{fig:histvol}, the bins on the left represent the smallest particle volumes, but contain the highest number of particles, indicated by a yellow color coding. As the volume increases from left to right in the visualization, the bins show a sharp decrease in particle counts (encoded in purple), with several bins on the far right containing no particles at all (encoded in black). This demonstrates the heavily tailed nature of the distribution. The varying shades of yellow and green in the leftmost bins allow for the detection of slight differences in particle counts, thereby providing an estimate of peak magnitude. Conversely, the purple bins on the right side reveal outliers with significantly larger volumes than the majority of the data. These outliers typically represent agglomerated particles that were not separated during segmentation. Furthermore, the spatial tiling utilized to generate the 54 distributions provides a coarse estimation of the regions containing these outliers. While UB binning offers a comprehensive overview of the distribution, it is unable to accurately resolve the peaks at the left boundary of the visualization, unless finer binning with a higher number of bins is applied. This subsequently compromises the clarity of the outliers in the other regions. 

\begin{figure}[tb]
 \centering
 \includegraphics[width=\linewidth]{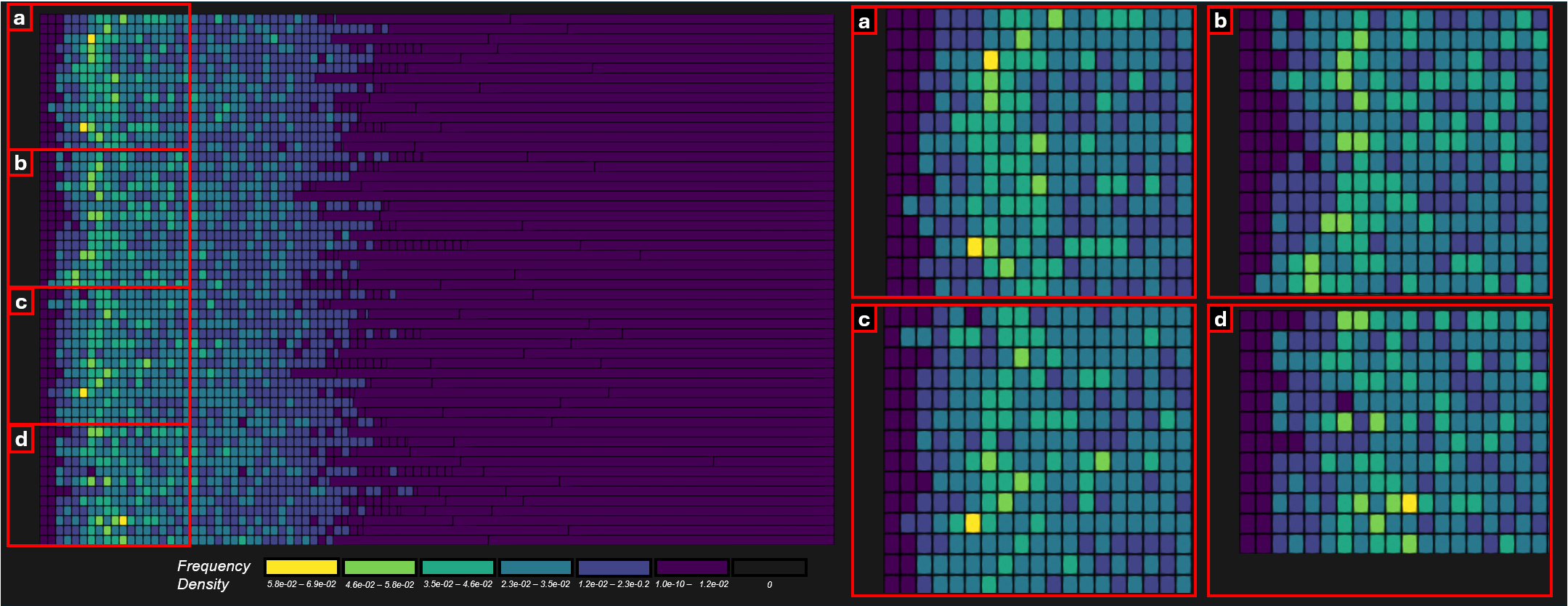}
 \caption{AccuStripes visualization of the particle volume characteristic employing BB and the \textit{color only} composition. The variation of the peaks is presented in greater detail over the spatial tiles, as illustrated in the detail visualizations (a)-(d). However, outliers are no longer identifiable}
 \label{fig:accustripesVolumeBB}
\end{figure}

To address these limitations, we used BB binning, as shown in Fig.~\ref{fig:accustripesVolumeBB}. This method constructs narrower bins around the peaks, while allocating wider bins for the less frequently occurring larger particles. This approach significantly improves the representation of the peaks on the left, which represent smaller particles, and provides a view of the pronounced tail of the distribution, similar to Fig.~\ref{fig:histvol}, but with much more detail in the regions of the peaks. The additional spatial information indicates that the distribution of particle volumes is relatively consistent throughout the sample. However, the detailed visualizations in Fig.~\ref{fig:accustripesVolumeBB} (a)-(d) demonstrate variations in particle volume across different spatial regions. Despite these advantages, BB binning shows limitations in identifying outliers, as they tend to be grouped with empty regions in large bins.
Overall, both considerations are essential to a comprehensive understanding of the distribution of particle volumes within the sample.

\medskip

\begin{figure}
 \centering
 \includegraphics[width=\linewidth]{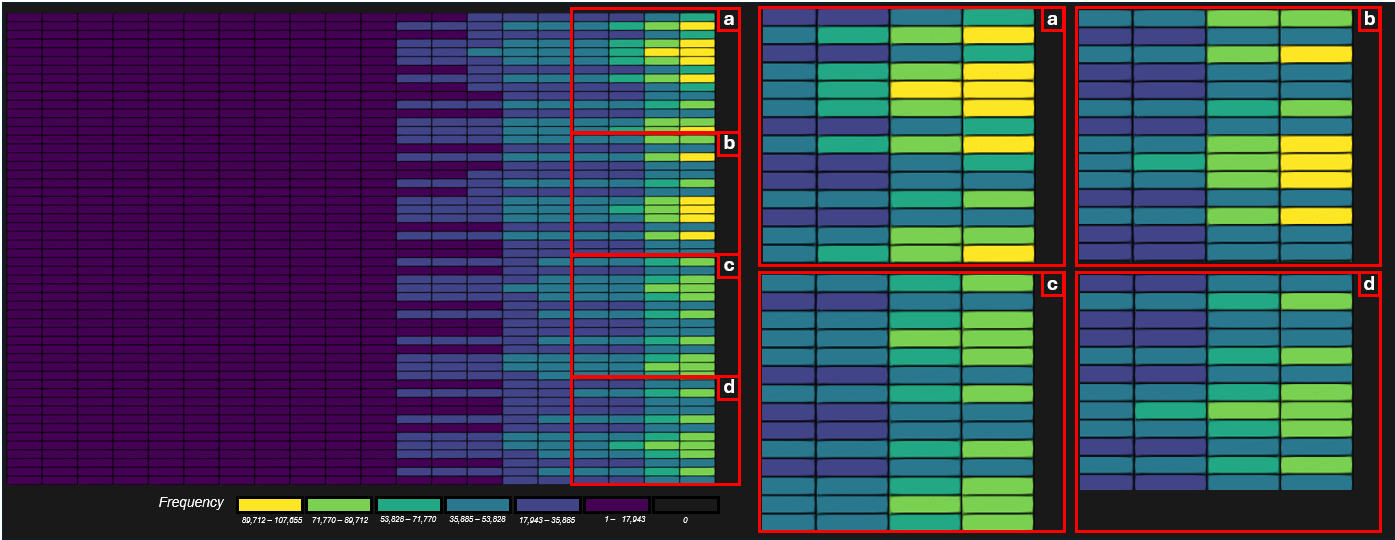}
 \caption{AccuStripes visualization of the particle sphericity property employing UB and the \textit{color only} composition. The smooth shape of the density function is depicted well, yet the spatial variation of particles of different sphericity values is shown in a misleading way, cf. the detail visualizations (a)-(d)}
 \label{fig:accustripesSphericityUB}
\end{figure}

A similar analysis is conducted for the sphericity property, which compares a particle to an ideal sphere and expresses this relation as a percentage, scaled to the interval $[0,100]$. Once more, considering UB with 20 bins yields the AccuStripes image shown in Fig.~\ref{fig:accustripesSphericityUB}. Similarly to the preceding analysis, the outcome is comparable to the global histogram illustrated in Fig.~\ref{fig:histroundness}, exhibiting a pronounced peak on the right-hand side of the visualization, corresponding to a high degree of sphericity.
In particular, the detailed visualizations shown in Fig.~\ref{fig:accustripesSphericityUB} (a)-(d) suggest that the "first half" of the sample, based on spatial tiling, appears to contain more particles with higher sphericity compared to the second half.

\begin{figure}
 \centering
 \includegraphics[width=\linewidth]{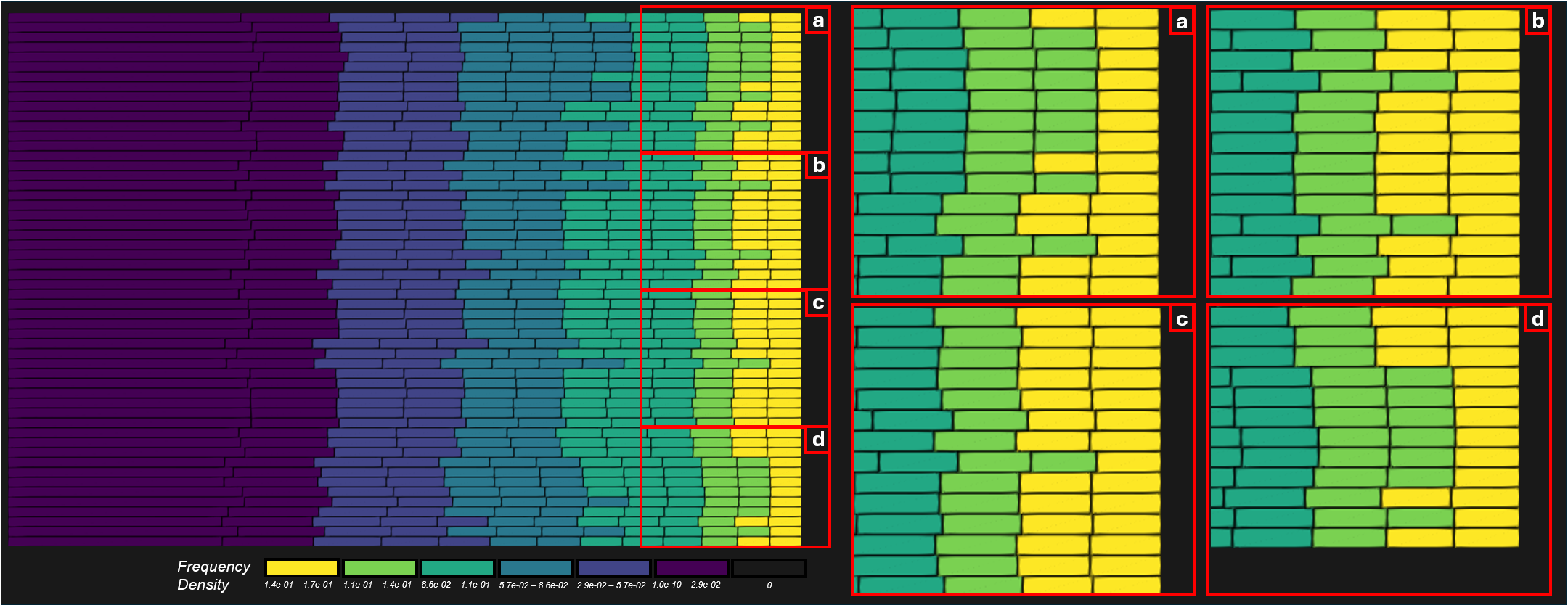}
 \caption{AccuStripes visualization of the particle sphericity property employing NB and the \textit{color only} composition. The shape of the distribution is well represented. Adaptive binning better accounts for small variations in sphericity values, cf. the detail visualizations (a)-(d)}
 \label{fig:accustripesSphericityNB}
\end{figure}

For an alternative perspective, NB binning is applied in Fig.~\ref{fig:accustripesSphericityNB}. The adaptive binning method reveals that all regions of the sample exhibit approximately similar distributions of the particles with regard to their sphericity. However, the UB binning method assigns particles with sphericity values just below a bin boundary to a different bin. This results in a distortion of the histograms, creating the false impression that only certain regions of the sample contain particles of a very round shape. In contrast, NB sets the bin boundaries adaptively, providing a clearer representation that complements with the initial observations from Fig.~\ref{fig:histroundness}.

As illustrated, AccuStripes may facilitate the comparison of multiple distributions across spatial regions, enabling users to identify patterns such as peaks or outliers, and to understand the variation of features such as particle volume or sphericity across a sample. Adaptive binning methods such as BB and NB can improve the interpretability of skewed data by adjusting bin widths based on the data distribution, providing finer detail around peaks.

\section{Limitations \& Future Work}
While AccuStripes effectively supports the comparison of distributions, it misses an inherent connection to the spatial context of the 3D volume. This limits the ability to localize certain features within the sample, such as identifying regions containing larger particles. A promising direction for future work is to integrate AccuStripes with 3D volume rendering, allowing users to interactively select bins in the AccuStripes representation and highlight the corresponding features in the spatial domain. This linkage would bridge the gap between abstract visualization and the spatial domain.

For extremely large high-resolution scans, resulting in hundreds of spatial regions and distributions, the stacked visualizations can become visually dense and difficult to interpret. To address this, abstraction techniques such as clustering of similar regions could be employed. Users could then explore aggregated summaries at a higher level and drill down into individual groups of regions for detailed inspection.
Finally, future research should explore progressive visual analytics strategies for AccuStripes to better handle large-scale scan data. Additionally, the methodology could be extended beyond particle-based applications to investigate its applicability in other domains of volumetric data analysis.

\section{Conclusion}\label{sec:conclusion}
This work demonstrates the effectiveness of advanced visualization techniques for detailed SCT data analysis on the example of applying AccuStripes to derived secondary data of a high-resolution scan of a novel material. AccuStripes facilitates an exploration of secondary derived data and enables users to examine the variability of particles' properties across the entire volume, with the majority of particles being relatively small and spherical. We demonstrated that the selection of binning methods has a significant impact on the interpretability of the data. While the use of UB provides a clear overview and can support the detection of outliers, adaptive techniques such as BB and NB can enhance the resolution of peaks and improve the representation of skewed distributions. Consequently, the employment of diverse binning techniques facilitates interpretation and avoids misleading visual representations. Nevertheless, we also identified limitations in conveying spatial context and handling even larger numbers of distributions, suggesting opportunities for future work. In conclusion, the integration of AccuStripes with adaptive binning provides an effective approach to uncovering structure in vast numbers of distributions and assists domain experts in interpreting secondary data extracted from primary volumetric data.


\backmatter
\bmhead{Acknowledgements}
We want to thank Paul Tafforeau (ESRF) for supporting the measurement process at the BM18 beamline.

\section*{Declarations}
\subsection*{Funding}
This work was supported by the German Federal Ministry of Education and Research (BMBF) within the project "KI4D4E: Ein KI-basiertes Framework f\"{u}r die Visualisierung und Auswertung der massiven Datenmengen der 4D Tomographie für Endanwender von Beamlines" under the title 05D2022 in a collaboration between the Fraunhofer Gesellschaft zur F\"{o}rderung der angewandten Forschung e.V., the University of Stuttgart, the University of Passau, the Friedrich-Alexander University of Erlangen, the Karlsruhe Institute of Technology, the Helmholtz center Berlin, the Helmholtz center Hereon, and the Forschungszentrum J\"{u}lich.

\subsection*{Competing interests}
The authors have no conflicts of interest to declare that are relevant to the content of this article.

\subsection*{Ethics approval and consent to participate}
Not applicable.

\subsection*{Consent for publication}
Not applicable.

\subsection*{Data availability}
%
The secondary data analyzed in this work may be shared upon reasonable request. 

\subsection*{Materials availability}
Not applicable.

\subsection*{Code availability}
Not applicable.

\subsection*{Author contribution}
All authors contributed to the conceptualization of the presented work. The funding acquisition and data curation was performed by Thomas Lang. Anja Heim and Thomas Lang developed the methodology. The implementation, data analysis, and visualization of the results using AccuStripes was done by Anja Heim. Thomas Lang participated in the result validation and Christoph Heinzl supervised the overall activities. Anja Heim and Thomas Lang contributed to writing the original draft, and all authors participated in review and editing of this work.


\bibliography{bibliography}

\end{document}